\documentclass[12pt]{article}

\usepackage{dcolumn}
\usepackage{bm}
\usepackage{graphicx}
\begin{document}

\title{Analysis of the cool down related cavity performance of the European XFEL vertical acceptance tests}
\author{M. Wenskat\footnote{marc.wenskat@desy.de}, J. Schaffran \\ Deutsches Elektronen Synchrotron, Hamburg, Germany}

\maketitle
\begin{abstract}
For the European X-Ray Free Electron Laser (XFEL) cavity production, the cold radio-frequency (RF) test of the cavities at 2\,K after delivery from the two vendors was the mandatory acceptance test. It has been previously reported, that the cool down dynamics of a cavity across $T_c$ has a significant influence on the observed intrinsic quality factor $Q_0$, which is a measure of the losses on the inner cavity surface.
A total number of 367 cool downs is used to analyze this correlation and we show that such a correlation is not observed during the European XFEL cavity production.
\end{abstract}
\section{Surface Resistance and Flux Trapping}
Superconducting cavities came a long way and are now an established technology for large scale accelerators. Still, for future particle accelerators, the optimization of the cavity performance is a relevant topic. The two most important parameters for the cavity operation
are the absolute value for the intrinsic quality factor $Q_0$ and its dependency on the applied accelerating gradient $\mathrm{E_{acc}}$, where the focus in this work will be on the quality factor.
The so called $Q_0$ versus $\mathrm{E_{acc}}$ curves are mandatory for the acceptance test of any large scale production. The European XFEL cavity production itself is described in detail elsewhere \cite{Singer, Reschke}. 
\newline
The quality factor $Q_0$ is inverse to the surface resistance of the cavity, with a geometry constant as a conversion factor. This surface resistance is usually described as
\begin{equation}
 R_s (T)= R_{BCS}(T) + R_0
\end{equation}
with $R_{BCS}(T)$ the temperature dependent contribution, described by the BCS-Theory and $R_0$ as residual resistance. Both contributions show a not fully understood dependency on the accelerating gradient $\mathrm{E_{acc}}$. This residual resistance is usually a ``melting pot'' for many different effects with varying dependencies.
The BCS term is a quite complex expression, which usually can be simplified by some approximations using the so called dirty limit case (mean free path $<$ coherence length), assuming a low frequency compared to the gap frequency and $T<T_c$. This yields to the well known equation of
\begin{equation}
 R_{BCS}(T)=\frac{A \omega^2}{T} \times \mathrm{exp}\left(-\frac{\Delta_0}{k_b T} \right)
\end{equation}
with $\Delta_0$ as the energy gap, $k_b$ the Boltzmann constant, $\omega$ the operating frequency, $T$ the operating temperature and $A$ a material constant. The ratio at 2\,K between $R_{BCS}$ and $R_0$ for the European XFEL cavity production is about 2:1 \cite{Reschke} and hence, a change of $R_0$ has a observable impact on the cavity performance.
As it has been reported in previous studies \cite{Vogt, Romanenko}, the cool down dynamics of the cavity around $T_c$ in an ambient magnetic field has a significant influence on the achievable quality factor. Models described elsewhere \cite{Kubo1, Kubo2, Huang, Checchin} relate the flux trapping resistance $R_{fl}$, as a major contribution to the residual resistance $R_0$, with either the cool down rate $\frac{dT}{dt}$ or spatial temperature gradient across the cavity $\frac{dT}{dx}$ while crossing $T_c$ as determining factor. 
As the material is cooled down, a phase transition front sweeps through the cavity. Magnetic vortices penetrating the cavity in the normal conducting phase are expelled when this transition front passes a pinning site (e.g. grain boundaries, dislocations, normal conducting and dielectric inclusions) 
if the thermal gradient locally exceeds the pinning force. Within these models, an inverse relation of $R_{fl}$ with the cool down rate $\frac{dT}{dt}$ or spatial temperature gradient across the cavity $\frac{dT}{dx}$ is obtained. 
Hence, a relation of the absolute value of the measured quality factor $Q_0$ and the cool down rate or spatial temperature gradient is expected.

\section{Experimental Set Up}
\begin{figure*}[!htp]
	\centering
		\includegraphics[width=8cm]{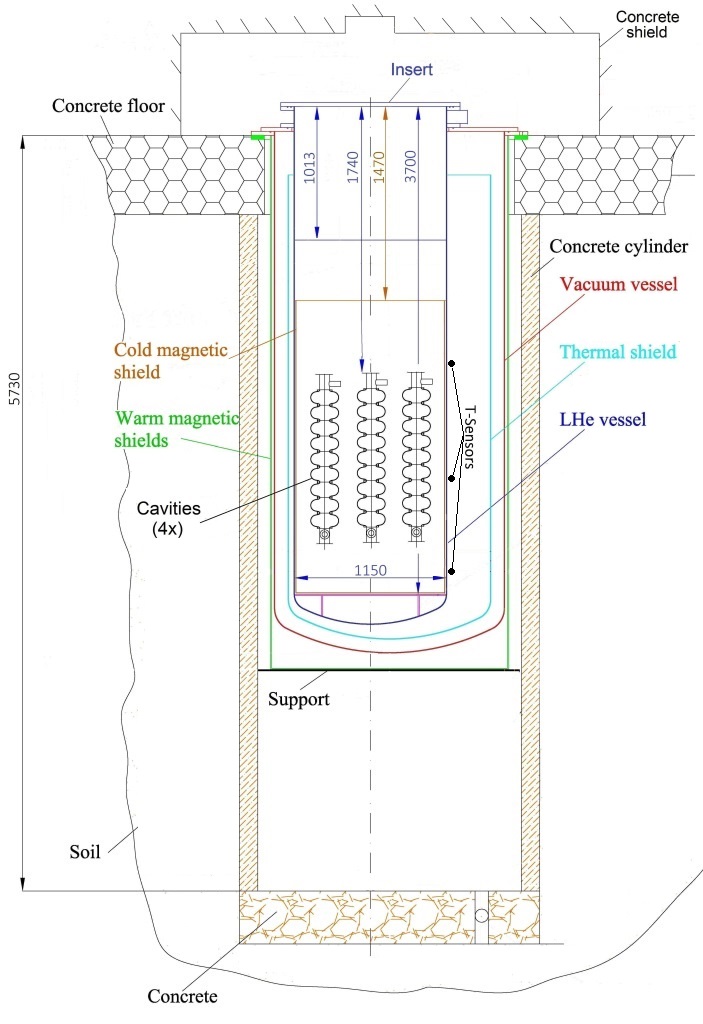}
	\caption{Section diagram of the vertical test cryostat. Note that only three of the four cavities are visible. The TVO sensor positions are indicated on the right side of the cryostat. One sensor is located at the vertical position corresponding at a location between equators three and four of the cavity; two other sensors were 880\,mm above and below this location.}
	\label{fig:Cryostatdesign}
\end{figure*}

A cross section of the vertical test stands\cite{Polinski}, called XATC1 and XATC2, at the Accelerator Module Test Facility\cite{Bozkov} (AMTF) can be seen in Figure \ref{fig:Cryostatdesign}. Each cryostat allows a simultaneous testing of 4 cavities mounted on a dedicated insert. The cool down procedure at the AMTF vertical test stands is mainly automated (PLC-based) and only requires operator supervision \cite{Petersen, Anashin}. Cernox\textsuperscript{TM} temperature sensors installed at one of the inserts were used to commission the automated procedures and monitor the thermal stress on the inserts \cite{Schaffran}. Carbon temperature sensors\cite{Datskov} (TVO) glued to the outside of the cryostat walls are used during normal operation. One sensor is located at a position corresponding to a location between equators three and four of the cavity; two other sensors are 880\,mm above and below that location. This instrumentation was designed for a well-defined operation and control of the cryogenic system and any cold vertical tests were performed in stable cryogenic conditions. During the vertical testing of the series production, neither the inserts nor the individual cavities were equipped with temperature sensors.  
\newline
\subsection{Cool Down Procedure}
\begin{figure}[!htp]
	\centering
		\includegraphics[width=8.8cm]{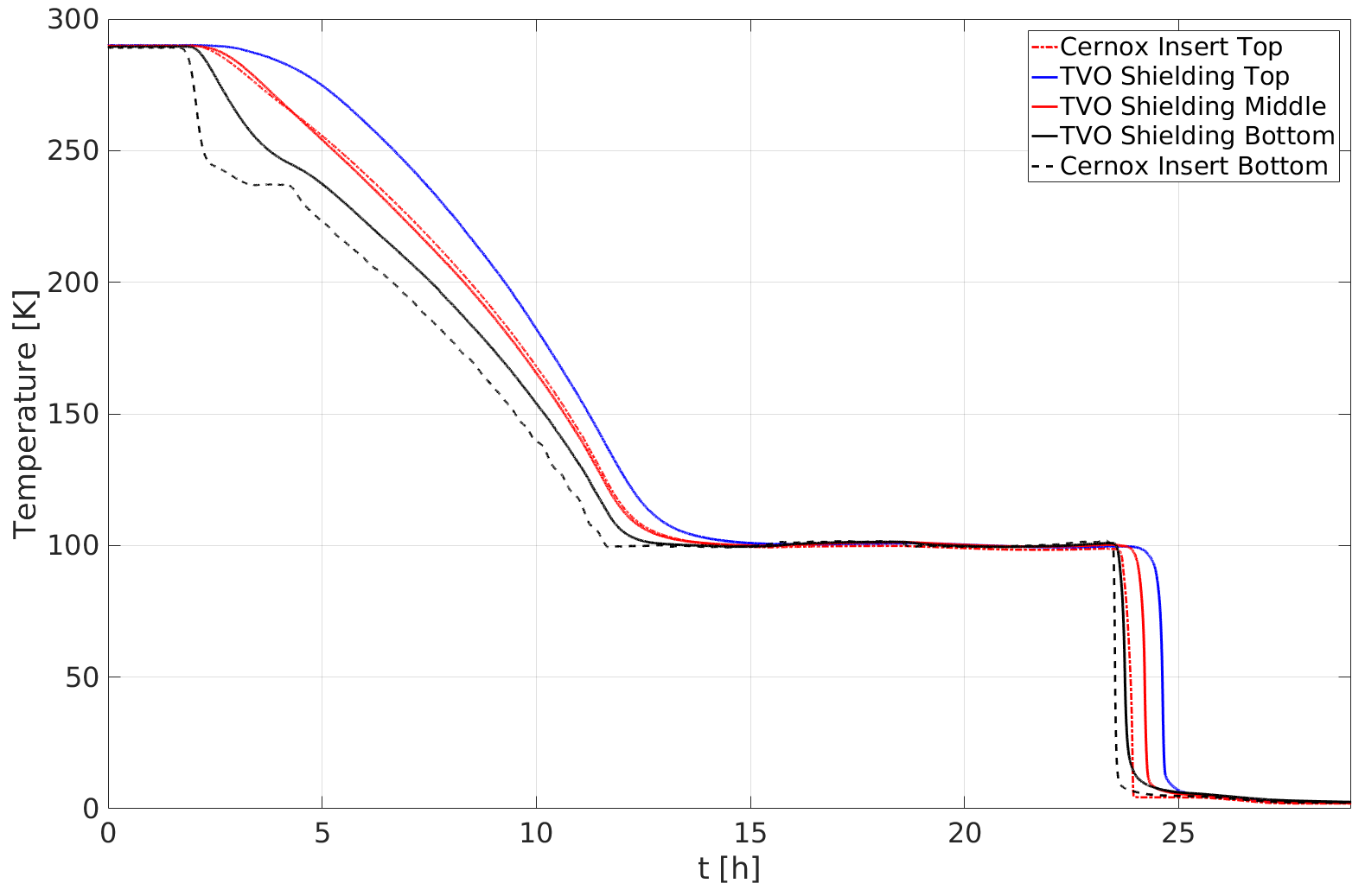}
	\caption{Temperature vs. time for three different TVO sensors glued to the outside of the cold shield and two Cernox\textsuperscript{TM} sensors, installed at the top and bottom of an insert.}
	\label{fig:Cooldowncycle}
\end{figure}

A full cool down is shown in Figure \ref{fig:Cooldowncycle}. The cool down from 300\,K to 100\,K took approximately 12\,hours, which represents an average cool down rate of 5\,mK/s. The cryostat remains at 100\,K for 4-6\,hours \footnote{The cavity is kept at 100\,K to foster the production of hydrides and hence observe the hydrogen Q-Disease during the acceptance test\cite{Hasan}. This prevents that such a deterioration remains undetected till after module assembly and tunnel installation.} after which the cavity is further cooled to a set point at 4\,K. The final cool down to 2\,K was performed manually
by reducing the vapor pressure of the helium bath to 30\,mbar, achieving an average cool-down rate of 0.5\,mK/s. The cool down rate across $T_c$ was up to 300\,mK/s at which point a maximal longitudinal temperature gradient along the cavities of up to 40\,K/m was observed. 

\subsection{DC magnetic field in the test cryostat}
\begin{figure*}[htp]
	\centering
		\includegraphics[width=1.00\textwidth]{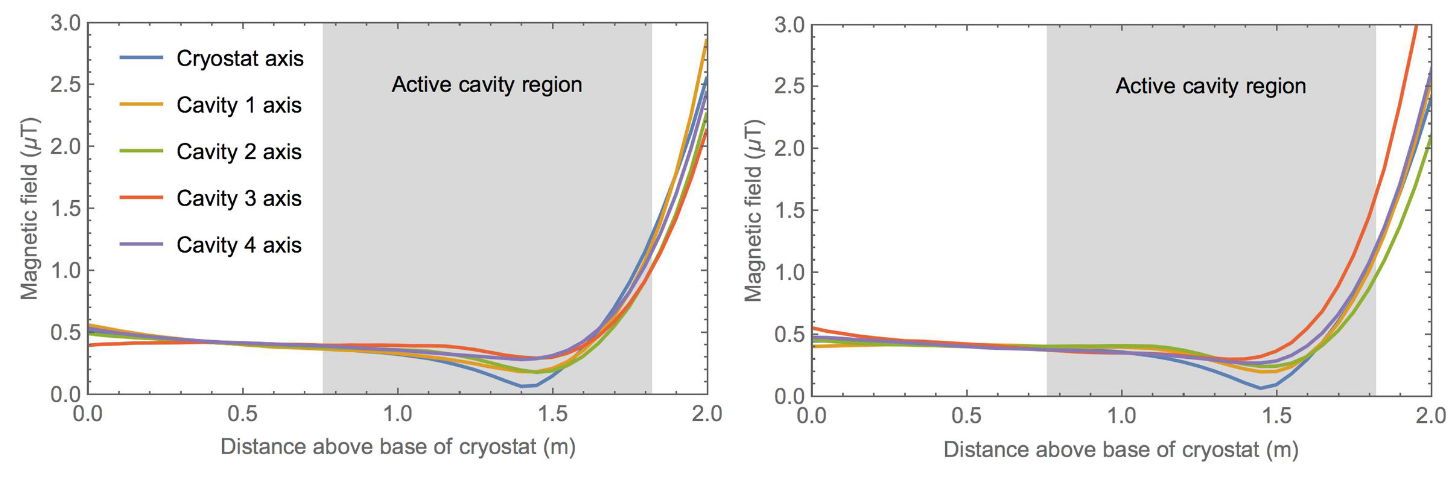}
	\caption{Magnetic field distribution inside the cryostat XATC1 (left) and XATC2 (right) along the different axes.}
	\label{fig:cryostat-B-measurements}
\end{figure*}
One significant influence on the quality factor $Q_0$ (or more specifically on the RF residual resistance $R_0$) is the ambient magnetic field during cool down. A suppression of the ambient magnetic field below a value of 1-2\,$\mu \mathrm{T}$ is necessary to achieve the European XFEL specifications for cavity performance\cite{Field1, Field2}. 
The earth magnetic field in the concrete pit for the two vertical cryostats was approximately 40\,$\mu \mathrm{T}$. Simulations showed that two magnetic shields were necessary: a double-walled warm magnetic shield made of mu-metal around the cryostat (300K), and an additional smaller one made out of Cryoperm\textsuperscript{TM} inside the cryostat in the 2\,K region. All shields were cylindrical with a closed bottom (see Figure \ref{fig:Cryostatdesign}). 
\newline
Measurements of the magnetic field inside of the closed cryostat were made at room temperature and are given in Figure \ref{fig:cryostat-B-measurements}. The magnetic field at 2\,K will be lower than at room temperature, since the Cryoperm\textsuperscript{TM} has a higher shielding efficiency at 2\,K than at room temperature.

\subsection{Comparison of TVO and Cernox\textsuperscript{TM} sensors}
\begin{figure}[!htp]
	\centering
		\includegraphics[width=8.8cm]{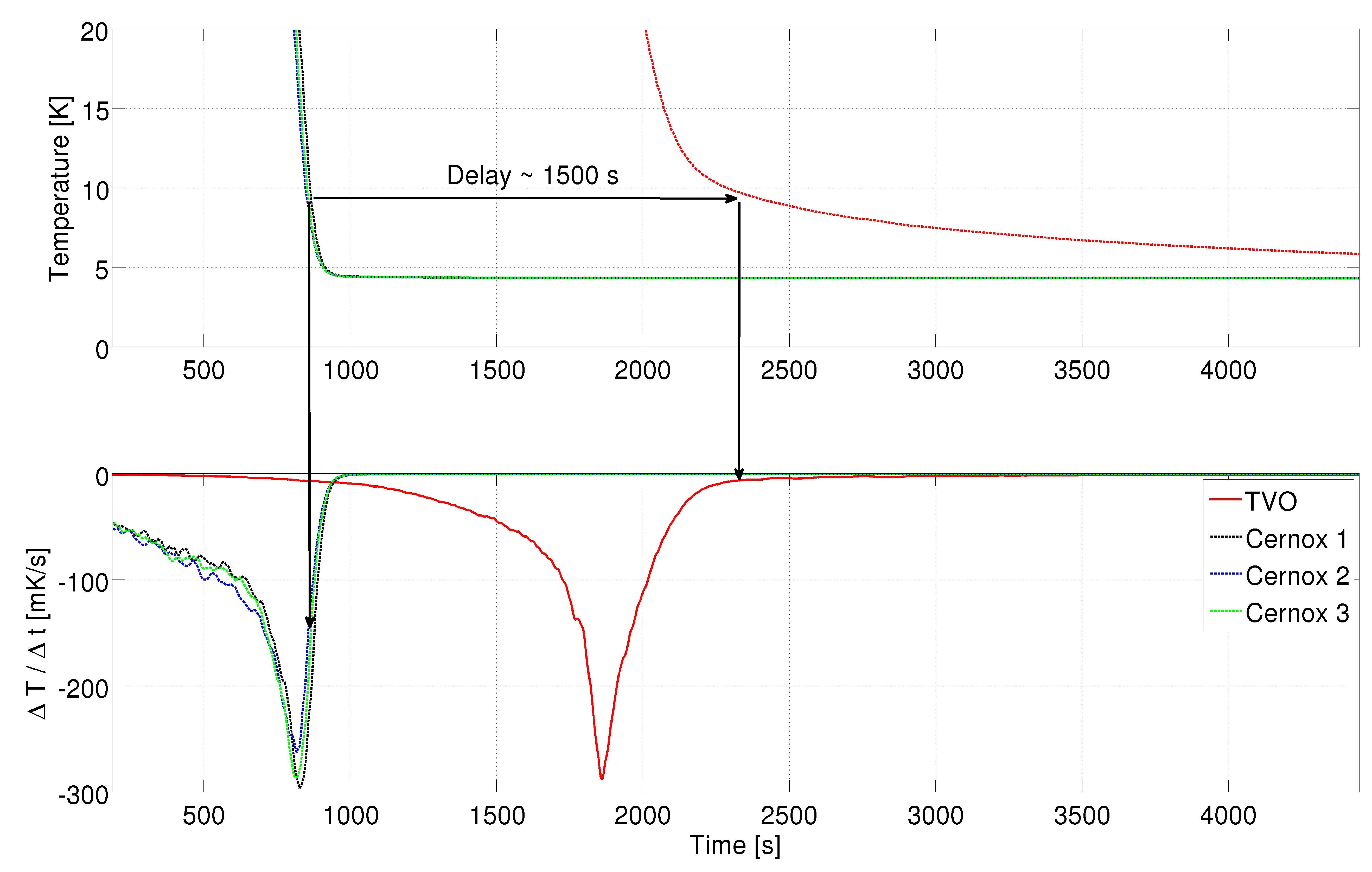}
	\caption{Top: Temperature against time for 3 Cernox\textsuperscript{TM} and 1 TVO sensor. A delay caused by the horizontal displacement of the sensors is visible (indicated by the horizontal arrow) and the TVO sensor shows a less steep behavior across $T_c$. Bottom: The cool down rate of the sensors against time. The rate across $T_c$ is indicated by two downward arrows. A difference of the cool down rate for the two sensor locations of two orders of magnitude is observed.}
	\label{fig:Cooldown_Curves}
\end{figure}

As only the TVO data were available for the production testing, an additional test to gain data using the insert fully-instrumented with Cernox\textsuperscript{TM} sensors was used to interpret the TVO data in terms of the cool down rates at the cavity.
Figure \ref{fig:Cooldown_Curves} shows a comparison of the observed temperature between the TVO sensor glued to the outside of the cryostat at the height of the iris between equator three and four and three Cernox\textsuperscript{TM} sensors attached at the insert and thermally connected to the cavity (1\,cm above and below the cavity). 
\newline
There is a delay of about 1500\,s in time at which $T_c$ is achieved between the Cernox\textsuperscript{TM} and TVO sensors, which is expected due to heat diffusion caused by the horizontal displacement of the sensors and the thermal capacity of the cryostat.
Hence, the to be compared cool down rate should not be identified by a given absolute time, but by a certain temperature achieved at the sensor, e.g. $T_c$. 
\newline
Figure \ref{fig:Cooldown_Curves} (bottom) shows the time derivatives (instantaneous cool down rate) of the sensor readouts. A much lower rate is observed at the TVO sensor at $T_c$ than with the corresponding Cernox\textsuperscript{TM} sensors.  For the measurement shown in Figure \ref{fig:Cooldown_Curves}, the cool down rate at the insert across $T_c$ was ~ 125\,mK/s while at the TVO sensors outside the helium tank the corresponding value was 5.4\,mK/s. This is a direct consequence of the above mentioned heat capacity of the helium tank. 
\newline
Using a simple thermal model, the effect of the heat capacity of the helium tank can be quantified. Using the differential heat diffusion equations, the temperature T at a sensor for a given time $t$ is $\propto exp\left(-t / c \right)$ with $c$ as a constant including the heat capacity of the material(s) between the heat source and the sensor. To estimate the ratio of the cool down rates between the TVO sensor at the height of equators three and four and the Cernox\textsuperscript{TM} sensor at the insert, the ratio of the first derivatives is used and the constants are obtained from a fit. Taking the diffusion time into account, a ratio of the cool down rate across $T_c$ about $20 \pm 12$ is expected while the observed ratio is 23.1. This allows to estimate the cool down rate at the cavity with the cool down rate obtained at the TVO sensor.
\newline
Summing up, the observed differences can be understood as a consequence of the test stand design and sensor positions. Since this set up didn't change during the European XFEL cavity production and testing, the underlying assumption is that the observed differences of the cool down rates of the different sensors at the different positions is constant for all the cool downs. Hence, although the TVO sensors underestimate the cool down rate at the cavity, this underestimation should be the same for all cool downs in this analysis and any correlation should be preserved.

\section{Results}
\begin{figure}[!htp]
	\centering
		\includegraphics[width=8.8cm]{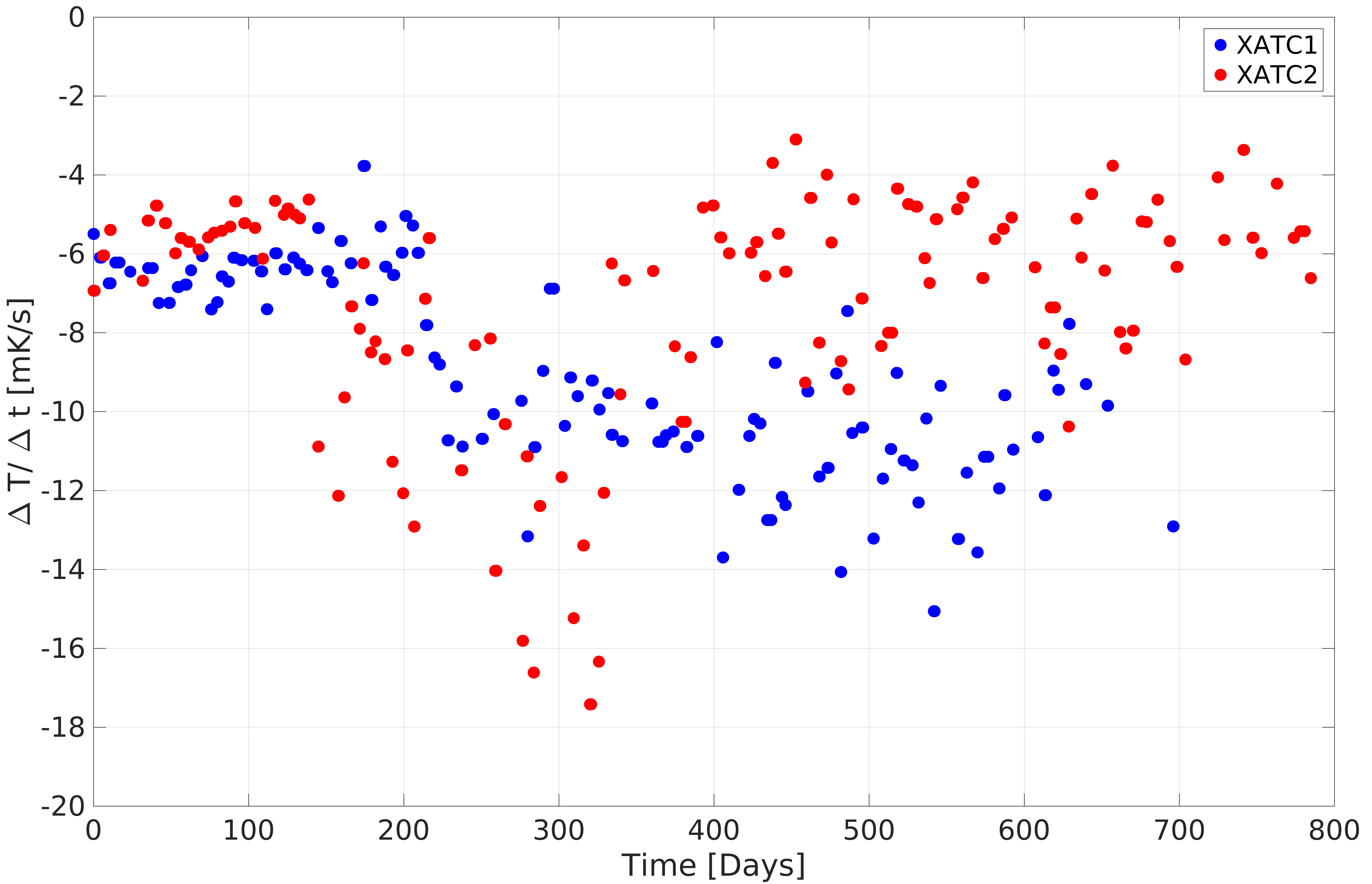}
	\caption{The horizontal axis shows the date when the cool down took place, relative to the beginning of the cavity production, while the vertical axis shows the cool down rate across $T_c$ for the respective cool down (obtained by TVO sensors glued to the outside of the cryostat). The operation of test stand XATC1 had to be stopped during the cavity production till the repair of a valve took place.}
	\label{fig:Cooldownrates_Temporal}
\end{figure}

\begin{figure}[!htp]
	\centering
		\includegraphics[width=8.8cm]{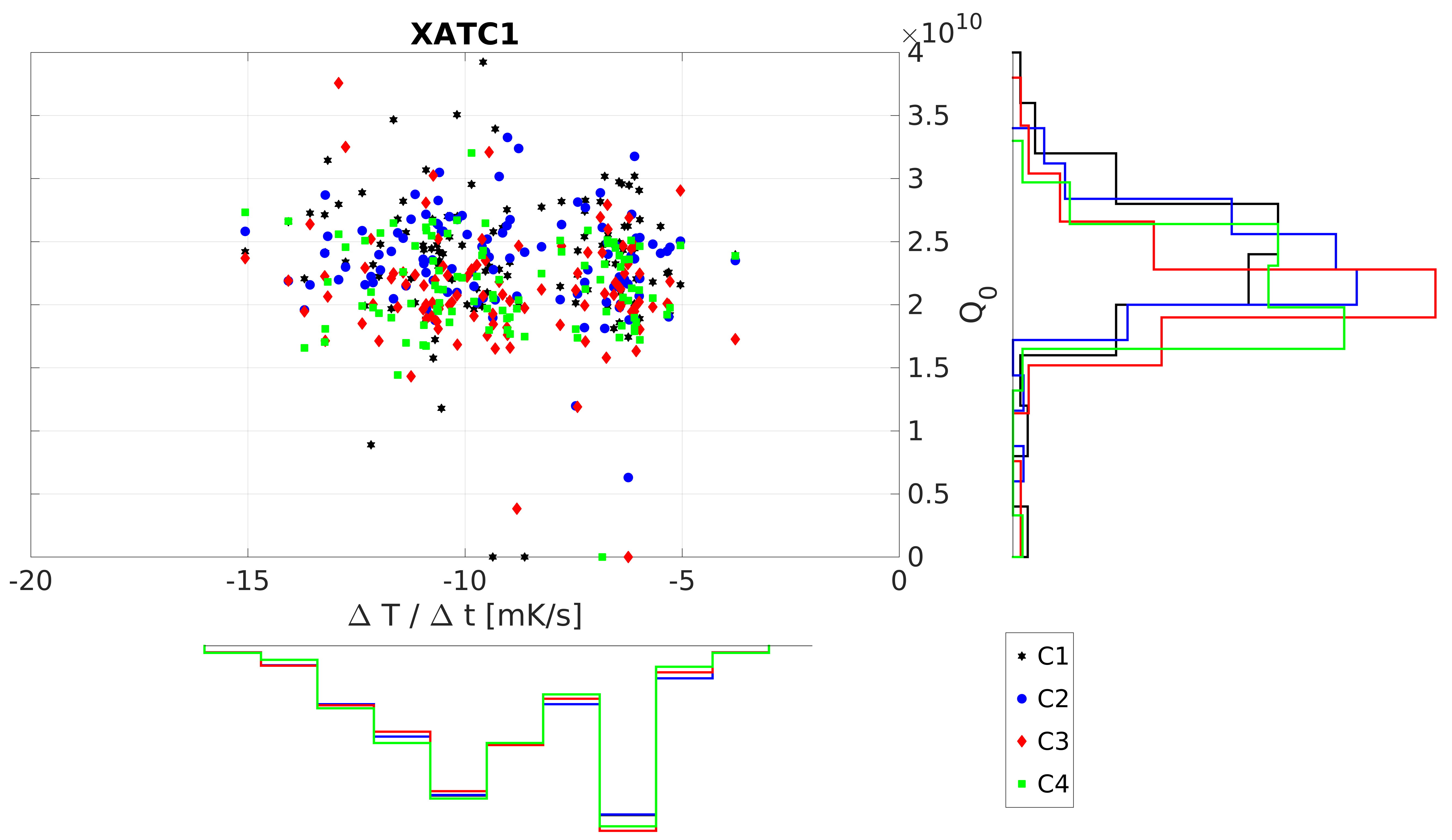}
	\caption{Scatterplot of $Q_0$ at 4\,MV/m measured in a vertical test versus the estimated cool down rate (obtained by TVO sensors glued to the outside of the cryostat), grouped by the four possible cavity positions C1-4 in vertical test stand XATC1. The histograms show the projected distributions.}
	\label{fig:Cooldown_V1_CX}
\end{figure}

\begin{figure}[!htp]
	\centering
		\includegraphics[width=8.8cm]{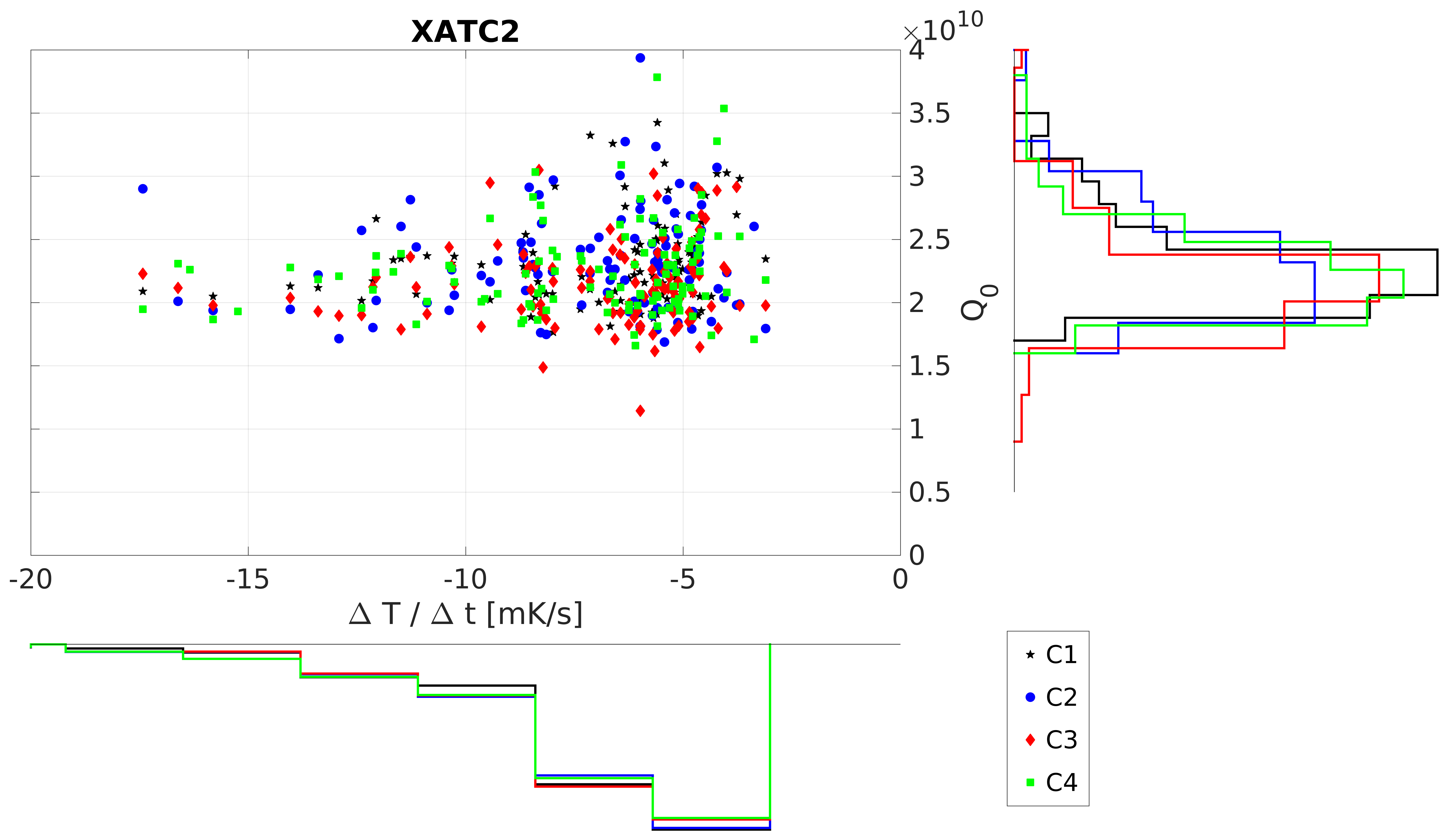}
	\caption{Scatterplot of $Q_0$ at 4\,MV/m measured in a vertical test versus the estimated cool down rate (obtained by TVO sensors glued to the outside of the cryostat), grouped by the four cavity positions in vertical test stand XATC2. The histograms show the projected distributions.}
	\label{fig:Cooldown_V2_CX}
\end{figure}

\begin{figure}[!htp]
	\centering
		\includegraphics[width=8.8cm]{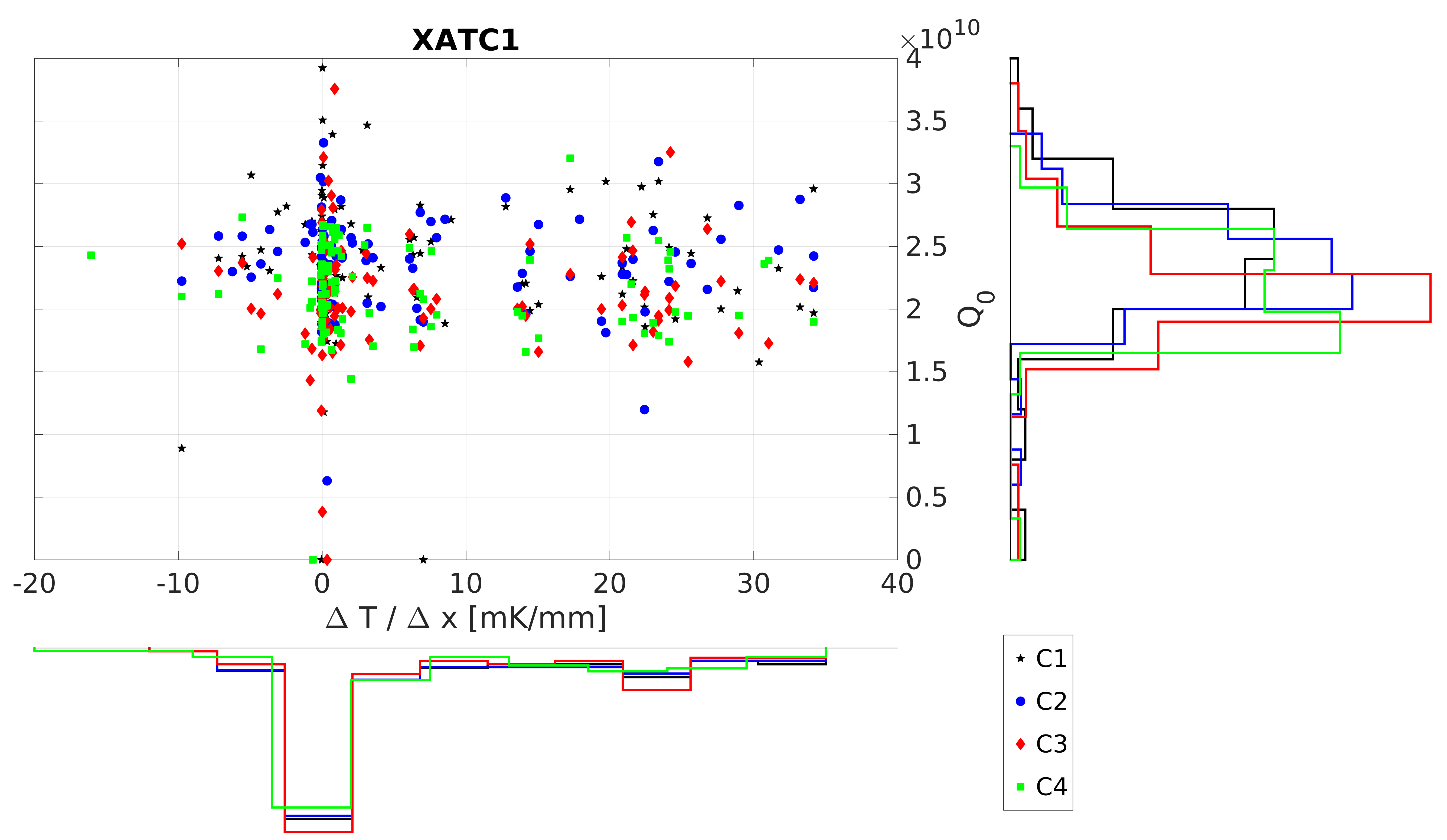}
	\caption{Scatterplot of $Q_0$ at 4\,MV/m measured in a vertical test versus the approximated spatial temperature gradient (obtained by TVO sensors glued to the outside of the cryostat), grouped by the four cavity positions in vertical test stand XATC1. The histograms show the projected distributions.}
	\label{fig:CooldownSpatial_V1_CX}
\end{figure}

\begin{figure}[!htp]
	\centering
		\includegraphics[width=8.8cm]{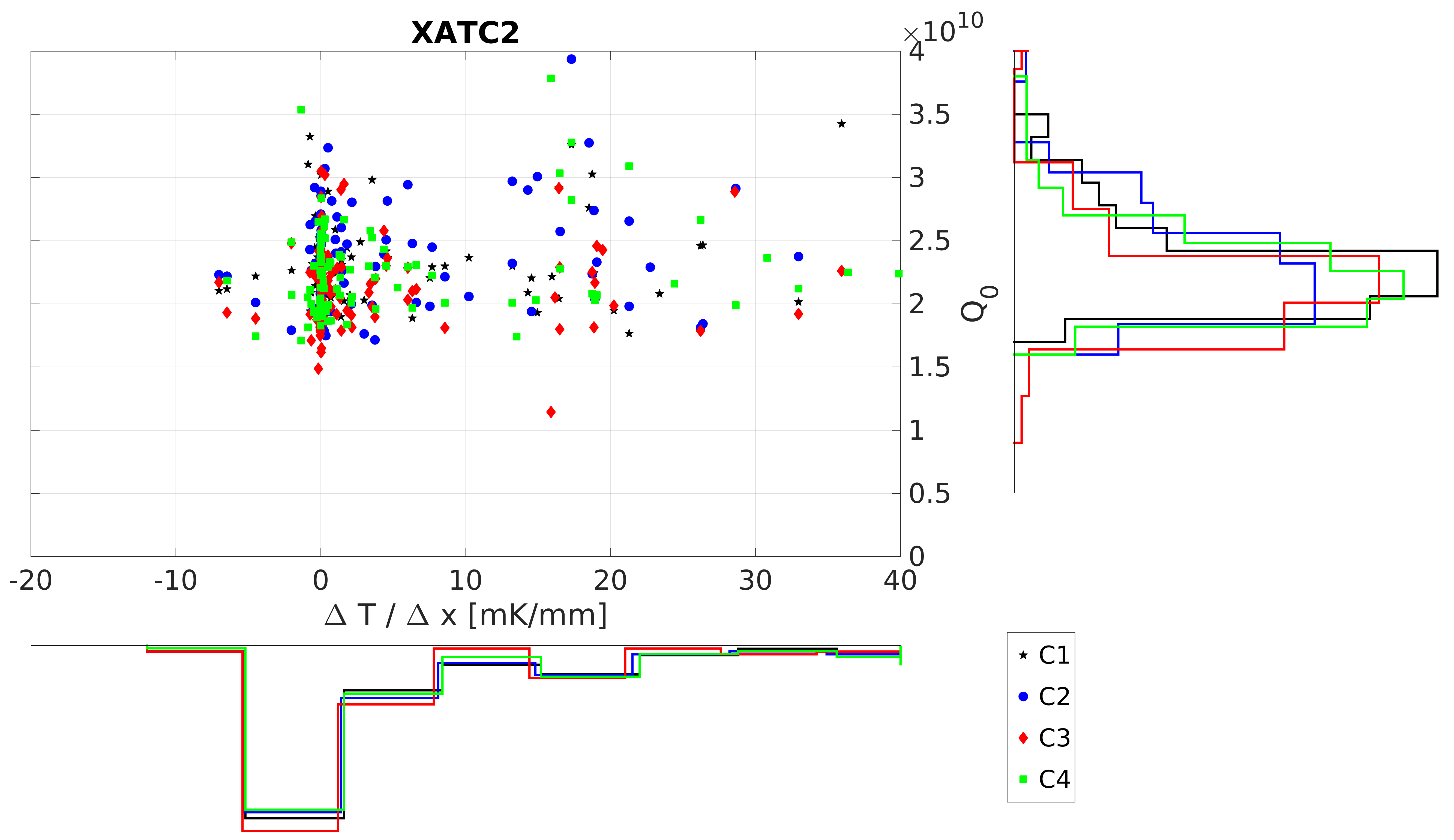}
	\caption{Scatterplot of $Q_0$ at 4\,MV/m measured in a vertical test versus the approximated spatial temperature gradient (obtained by TVO sensors glued to the outside of the cryostat), grouped by the four cavity positions in vertical test stand XATC2. The histograms show the projected distributions.}
	\label{fig:CooldownSpatial_V2_CX} 
\end{figure}

Since up to four cavities can be installed in one insert, groups of cavities tested together will have the same cool down rate. Cool down data were available for a total number of 168 (199) cool downs for test stand XATC1 (test stand XATC2), which results in a total of 457 (516) individual cavity tests recorded with the TVO sensors outside of the cryostat. Figure \ref{fig:Cooldownrates_Temporal} shows the temporal development of the cool down rates across $T_c$ for the two cryostats used. 
The use of cryostat XATC1 has to be stopped due to a blocked valve which prohibited any further operation, hence no more data points are available after approx. two years. The uniform behavior at the beginning of the production (up till day 200), was to ensure mechanical stability of the inserts. The experience gained after this first operational period showed that the mechanical stress due to cool down is less than expected and a faster cool down was possible resulting in a larger spread of cool down rates. 
\newline
Figures \ref{fig:Cooldown_V1_CX} and \ref{fig:Cooldown_V2_CX} show the scatter plot of the measured $Q_0$ at a accelerating gradient of 4\,MV/m in a vertical test versus the cool down rate across $T_c$, grouped according to cavity location in the test stand. This accelerating gradient value is chosen because of two reasons. First, because of a local maximum of the quality factor in every Q vs. $\mathrm{E_{acc}}$ curve at this value and serves as a standard. Second, because at that low accelerating field, no dominant loss mechanism limiting the cavity is present which would falsify the possible influence of the cool down dynamics on the quality factor.
The first observation is that no correlation was found. There is also no correlation observed, if one looks for the subgroups of cavities mady out of the three different niobium supplier materials or for the two different final surface treatments, where the plots are not shown here. 
The second observation is the double Gaussian distribution of the cool down rate for the vertical test stand XATC1 (Figure \ref{fig:Cooldown_V1_CX}) and a large tail to higher values in the cool down rate for vertical test stand XATC2 (Figure \ref{fig:Cooldown_V2_CX}). This corresponds to the temporal distribution of the cool down rates as shown in Figure \ref{fig:Cooldownrates_Temporal}.
\newline
As suggested elsewhere \cite{Kubo1, Kubo2}, the spatial temperature gradient across the cavity during cool down may be the more precise parameter to model the influence of the cool down dynamics on the cavity performance. Hence, a second analysis considering this parameter was performed. The spatial gradient was approximated by taking the temperature difference of the upper and lower TVO sensor of the cryostat when the lower TVO sensor measured $T_c$ (in contrast for the cool down rate, where the middle sensor was utilized). The Figures \ref{fig:CooldownSpatial_V1_CX} and \ref{fig:CooldownSpatial_V2_CX} show the result of this analysis. 
Again, no correlation is observed. The negative values of the spatial gradient occurred when the filling \& cooling of the cryostat has been done from top to bottom. This was only done in a few cases and were caused by technical reasons. Extreme values of the spatial temperature gradient correspond to high cool down rates, which show the consistency of this analysis.   
\section{Discussion}
The European XFEL cavity production allowed a statistical significant analysis on several topics influencing the cavity performance. One interesting topic is the influence of the cool down dynamics across $T_c$, which was thoroughly investigated in this work. Given the data of 367 cool downs, no correlation between the cool down dynamics and the observed quality factor $Q_0$ was found. 
\newline
The fact that no correlation between the cool down rate or the spatial temperature gradient and the quality factor $Q_0$ is observed is not necessarily in contradiction to the observations made elsewhere. Two reasons can be the cause of this. First, it may be simply a consequence of the choice of the diagnostics system and the test stand geometry. The presented diagnostic design is sufficient for the acceptance tests performed for the European XFEL cavity production, since any measurements were done after a certain latency to give the whole system time to achieve a thermal equilibrium. 
This process is well monitored with the given diagnostic system. The observation of any correlation of such a kind as investigated here may be prevented in the AMTF hall, since such a time resolved measurement was not needed and hence not included in the specifications of the diagnostic system design. Especially the spatially temperature gradient $\frac{dT}{dx}|_{T=T_c}$ can be falsified with the given sensor placement and heat capacity of the helium tank. 
\newline
A second possible interpretation of the non-observable correlation is, that the ambient magnetic field of 1-2 $\mu \mathrm{T}$ is simply to weak to excite strong vortices and reduce the quality factor significantly for European XFEL cavities at the observed cool down gradients. This interpretation is supported by the findings in other studies \cite{Romanenko, Mattia2017}, which state that cool down rates above 30\,mK/s across $T_c$ at the cavities (corresponds to 1.3-1.5\,mK/s at the TVO sensors) lead to the theoretical maximum for flux expulsion.
Since all cool down rates for the European XFEL cavities seem to be above this value when extrapolating the TVO data onto the cavity position using the Cernox\textsuperscript{TM} comparison, the maximum amount of flux was already expelled and no further dependency should have been observed.


\section{Acknowledgments}
Thanks to the IFJ-PAN Krakow team which operated the AMTF hall over the cavity production period, Markus M\"oller for the technical details concerning the database and Yuri Bozhko concerning details on the sensors and the XFEL-DESY team which provided the data.

\end{document}